\documentstyle[aps,prl]{revtex}

\def\eeq{\end{equation}}
\def\beq{\begin{equation}}
\def\bea{\begin{eqnarray}}
\def\eea{\end{eqnarray}}
\begin{document}
\title{Two-dimensional maps at the edge of chaos: \\Numerical results 
for the Henon map}
\author{Ugur Tirnakli}
\address{Department of Physics, Faculty of Science, Ege University,
35100 Izmir, Turkey \\
tirnakli@sci.ege.edu.tr}
\maketitle

\begin{abstract}
The mixing properties (or sensitivity to initial conditions) of 
two-dimensional Henon map have been explored numerically at the edge of chaos. 
Three independent methods, which have been developed and used so far for the 
one-dimensional maps, have been used to accomplish this task. These methods 
are (i)~measure of the divergence of initially nearby orbits, (ii)~analysis 
of the multifractal spectrum and (iii)~computation of nonextensive entropy 
increase rates. The obtained results strongly agree with those of the 
one-dimensional cases and constitute the first verification of this scenario 
in two-dimensional maps. This obviously makes the idea of weak chaos even 
more robust.\\

\noindent
{\it PACS Number(s): 05.45.-a, 05.20.-y, 05.70.Ce}
\end{abstract}



\vspace{1.5cm}

Nowadays a growing number of works addressing one-dimensional dissipative 
maps signals out the increasing interest in exploring the behaviour of 
nonlinear dynamical systems at the edge of 
chaos\cite{TPZ,costa,lyra,circle,latbar2,ugur,ugur2,ugur3,ugur4}. 
This interest mostly stems from the fact that the standard theory considers 
this special point (and also all other points where the standard Lyapunov 
exponent vanishes) as {\it marginal} and works addressing directly these 
marginal points are less than the works on the other regions of the 
system. Recent works addressing the behaviour of dynamical 
systems at these points are based on the conjecture that the divergence of 
initially nearby trajectories, characterized by the sensitivity function 

\beq
\xi(t)=\lim_{\Delta x(0)\rightarrow 0} \frac{\Delta x(t)}{\Delta x(0)}
\eeq	
(where $\Delta x(0)$ and $\Delta x(t)$ are the discrepancies of the initial 
conditions at times 0 and $t$), are not of the exponential type at these 
marginal points, but rather, of the power-law type like \cite{costa}

\beq
\xi(t)= \left[1+(1-q)\lambda_q t\right]^{1/(1-q)}\;\;\;\;\;\;
(q\in {\cal R}),
\eeq
which comes from the conjecture that the controlling equation becomes 
$d\xi / dt=\lambda_q \xi^q$ (instead of the usual one $d\xi/dt=\lambda_1\xi$). 
This result recovers the standard one in the $q\rightarrow 1$ limit and also 
defines a generalized version of the Lyapunov exponent $\lambda_q$, 
which inversely scales with time, but now within a power-law. 
Within this unified scheme, apart from the standard regimes where we have 
chaos, sensitivity and insensitivity to the initial conditions, we have 
also weak insensitivity to the initial conditions for 
$q>1$, $\lambda_1=0$, $\lambda_q<0$ and weak sensitivity to the initial 
conditions for $q<1$, $\lambda_1=0$, $\lambda_q>0$. 
The last case characterizes also the most interesting marginal point, the 
chaos threshold. This kind of asymptotic power-law sensitivity to the 
initial conditions has already been observed since 
long\cite{grass,politi,mori}, but Eq.(2) provides a more complete description 
than that, since it is expected to be valid not only at very large times 
but also at intermediate times as well (of course after a short transient). 
This equation in fact corresponds to the power-law growth of the upper bounds 
of a complex time dependence of the sensitivity function, which includes 
considerable and ever lasting fluctuations. These upper bounds 
($\xi(t) \propto t^{1/(1-q)}$) allow us to determine the proper value of the 
$q$ index (say $q^*$) for the dynamical system under consideration. This 
constitutes the first method of finding the $q^*$ value of a given dynamical 
system.

The second method for the same purpose is based on the geometrical aspects of 
the critical attractor of the system at the edge of chaos. 
Due to the complexity of critical dynamical attractor, a multifractal formalism 
is needed in order to reveal its complete scaling behaviour. In this formalism, 
each moment of the probability distribution has a contribution coming from 
a particular subset of points of the attractor with fractal dimension $f$. 
Using a partition containing $M$ boxes, the content on each contributing box 
scales as $M^{-\alpha}$. The multifractal measure is then characterized by 
the continuous function $f(\alpha)$ which reflects the fractal dimension of 
the subset with singularity strength $\alpha$\cite{multif1,multif2}.
Using the scaling behaviour of the multifractal singularity spectrum 
$f(\alpha)$, a new scaling relation has been proposed \cite{lyra} as 

\beq
\frac{1}{1-q^*} = \frac{1}{\alpha_{min}} - \frac{1}{\alpha_{max}}\;\; 
\;\;\;\;\;\;\;\;\;  (q^* < 1)
\eeq
where $\alpha_{min}$ and $\alpha_{max}$ are the vanishing points of downward 
parabolalike concave curve $f(\alpha)$ and characterize the scaling behaviour 
of the most concentrated and most rarefied regions on the attractor. This 
fascinating relation, which connects the power-law sensitivity of dynamical 
systems with purely geometrical quantities, establishes another independent 
method for estimating the $q^*$ value of a dynamical system under study.

Finally, the third method of accomplishing the same task, deals with the 
entropy increase rates \cite{latbar2,ugur}. For large classes of dynamical 
systems, the rate of information loss in time is characterized by the 
Kolmogorov-Sinai (KS) entropy $K_1$, which is defined as the increase, per 
unit time, of the standard Boltzmann-Gibbs entropy 
$S_1=-\sum^W_{i=1} p_i \ln p_i$, namely 

\beq
K_1\equiv \lim_{t\rightarrow\infty}\lim_{W\rightarrow\infty}
\lim_{N\rightarrow\infty} \frac{S_1(t)}{t} \; 
\eeq
where $t$ is the time, $W$ is the number of regions in the partition of the 
phase space, $N$ is the number of initial conditions (all chosen at $t=0$ 
within one region among the $W$ available ones) that are evolving in time. 
Although the KS entropy is defined, in principle, in terms of a single 
trajectory in phase space, it appears that, in almost all cases, it is 
possible to replace it by one based on an ensemble of initial conditions. 
The third method basically depends on the ensemble version of generalized 
KS entropy $K_q$, which has been introduced as the increase of a proper 
nonextensive entropic form, namely, 

\beq
K_q\equiv \lim_{t\rightarrow\infty}\lim_{W\rightarrow\infty}
\lim_{N\rightarrow\infty} \frac{S_q(t)}{t} \; 
\eeq
where $S_q$ is the Tsallis entropy defined as \cite{ts1}

\beq
S_q(t)= \frac{1-\sum_{i=1}^{W} [p_i(t)]^q}{q-1}\; 
\eeq
where $W$ is total number of configurations and $\{p_i\}$ are the associated 
probabilities. 
It is clear that, consistently, the well-known Pesin equality is also expected 
to be generalizable as $K_q=\lambda_q$ if $\lambda_q > 0$ and $K_q=0$ 
otherwise. Within this scheme, it is conjectured that (i)~a special $q^*$ 
value exists such that $K_q$ is finite for $q=q^*$, vanishes for $q>q^*$ and 
diverges for $q<q^*$; (ii)~this special $q^*$ value coincides with that coming 
from the two other distinct methods explained above. This also connects $q^*$ 
values with the entropic index $q$ of the Tsallis entropy as mentioned 
in \cite{ramanugur}. 
At this point, it is worthwhile to clarify a subtlety related to the generalized 
Pesin equality: In fact, the $q$-generalized Pesin equality implies two things. 
Firstly, the values of $q$ determined through the entropy rates are the same 
as those measured through other two distinct methods; secondly, the slopes of 
the entropy rates are the same as the generalized Lyapunov exponents of the 
sensitivity to the initial conditions. Throughout the present paper, we address 
and verify only the first point, not the second.

These three independent methods of obtaining $q^*$ values have already been 
tested and verified with numerical calculations for a variety of 
one-dimensional dissipative (symmetric and asymmetric) map 
families \cite{TPZ,costa,lyra,circle,latbar2,ugur,ugur2,ugur3,ugur4}, which 
strongly supports the idea that all these methods yield one and the same 
proper $q^*$ value of a given map. On the other hand, up to now (to the best 
of our knowledge), there exist no attempt of testing these three methods in 
more general grounds like (i)~two-(or more) dimensional dissipative maps, 
(ii)~conservative maps, (iii)~high dimensional dissipative and conservative 
systems. In this study, our aim is to analyze a two-dimensional dissipative 
map within these lines for the first time, which hopefully constitutes an 
important step forward as, we believe, it would presumably stimulate new 
efforts addressing the other general cases given above. 

The two-dimensional map that we focus on here is the Henon map \cite{henon}

\begin{eqnarray}
F(x,y): \left(\begin{array}{cc}
x  \\ y 
\end{array} \right)\rightarrow 
\left(\begin{array}{cc}
1-ax^2+y \\ bx
\end{array} \right)
\end{eqnarray}
where $a$ and $b$ are map parameters. This map reduces to the standard logistic 
map when $b=0$, whereas it becomes conservative when $b=1$, in between these 
two cases, it is a two-dimensional dissipative map. Specifically, we focus on 
small values of the $b$ parameter such as $b=0.001$, $0.01$, $0.1$, etc 
since, for greater values of $b$, the numerical procedures used in the first 
and second methods do not allow us to analyze the map properly due to the 
increasing complexity of the system. To calculate the largest Lyapunov 
exponent of the system, among the numerical procedures like 
Benettin et al \cite{benettin} or Froyland \cite{froyland} algorithms, 
we use the procedure given in \cite{aziz}.

Before introducing our numerical findings of the three methods discussed above, 
let us recall our expectations. 
Since the Henon map, for all $b$ values, belongs to the same 
universality class of the logistic map (namely, it has the same Feigenbaum 
numbers and box counting fractal dimension at the edge of chaos with the 
logistic map, irrespective of the $b$ values, although the chaos threshold 
value $a_c$ monotonically decreases for increasing $b$ values), the natural 
expectation is to find a $q^*$ value for the Henon map that coincides with the 
value of the logistic case ($q^*\simeq 0.24$) for all $b$ values. 
However, since the verification of these three methods (employed so far to 
study only one-dimensional maps) in higher dimensional maps is quite valuable, 
it is better to use a test system with well known critical behaviour such as 
the Henon map. Within this reasoning, the present work will hopefully bring 
relevant information about the proper procedure to be used in investigating 
other high dimensional systems belonging to distinct universality classes.\\

\noindent {\bf First Method -} ~ Using the numerical procedure given 
in \cite{aziz} for the calculation of the Lyapunov exponent, one can determine 
the time evolution of the sensitivity function. The procedure is the following: 
Using the derivative of the Henon map $D F(x,y)$, one can compute how an 
infinitesimally small error in a point ($x,y$) of the attractor is transformed 
by one iteration. For any arbitrary direction ($\cos(\phi),\sin(\phi)$) of 
the error, one can compute the transformed error using the equation 

\begin{eqnarray}
D F(x,y) \left(\begin{array}{cc}
\cos(\phi)  \\ \sin(\phi) 
\end{array} \right) = 
\left(\begin{array}{cc}
-2ax\cos(\phi)+\sin(\phi) \\ b \cos(\phi)
\end{array} \right)\; .
\end{eqnarray}
Therefore the error amplification is measured by the factor 
$[(-2ax\cos(\phi)+\sin(\phi))^2+(b\cos(\phi))^2]^{1/2}$. 
By iteration and renormalizing this procedure one can obtain the largest 
Lyapunov exponent.
As it is seen in Fig.~1, it exhibits a power-law divergence 
($\xi \propto t^{1/(1-q^*)}$) at the edge of chaos and the upper bound slope 
of this fractal structure allows us to calculate the $q^*$ value. 
For the ($x_0,y_0$) pairs, we use the numerically determined extremum values 
of $x$ and its corresponding $y$ values for each $b$ parameter. For example, 
$(x_0,y_0)=(1,0)$ for $b=0$ (logistic case), 
$(x_0,y_0)=(1.0008451...,-0.0000001749...)$ for $b=0.001$ and 
$(x_0,y_0)=(1.0084814...,-0.0000198735...)$ for $b=0.01$. 
The values of critical $a$ parameter calculated at the edge of chaos for 
some $b$ values are the following: $a_c=1.40115518...$ for $b=0$, 
$a_c=1.39966671...$ for $b=0.01$ and $a_c=1.38637288...$ for $b=0.01$. 
From the analysis of Fig.~1, although the fractal structure and the slope 
seem to deteriorate with increasing values of $b$, this does not prevent us 
to conclude that the proper $q^*$ value of the Henon map, for all $b$ values, 
is $q^*\simeq 0.24$ since it is clearly seen that the fractal structure and 
the slope coincide with the logistic case ($b=0$) for smaller time steps 
(in logarithmic scale) as $b$ values increase, whereas they start to 
deteriorate for larger time step. As the values of $b$ increase, the beginning 
of this deterioration shifts to smaller time steps. This originates from the 
fact that one needs $a_c$ values with greater precision (in our calculations 
we have 10-11 digit precision) as $b$ values increase, since the map becomes 
to be more complex. More precisely, this effect is the same as the one that 
we can face with even for the logistic map if lower precision is used for 
$a_c$ value (for example, instead of 10-11 digit precision if one uses 6-7 
digits for $a_c$, then a similar type of deterioration in the fractal 
structure is seen for smaller time steps for the logistic case as well). 
This point will become more transparent below when we discuss the second 
method in obtaining the proper $q^*$ value of the Henon map.\\

\noindent {\bf Second Method -} ~ Now we turn our attention to the 
multifractal singularity spectrum $f(\alpha)$ of the critical attractor. 
If we can construct the $f(\alpha)$ curve for the Henon map, then it is 
evident from the scaling relation (3) that, using the end points of the 
spectrum, we can determine the proper $q^*$ value of this map.
Constructing the $f(\alpha)$ curve, we use the well-known procedure of 
Halsey et al. \cite{multif1}, however it is suitable here to mention that 
other algorithms such as Cvitanovic et al. \cite{cvi} can also be used for 
this purpose.  
Before giving our results for the Henon map, we believe, it is better to 
reexamine the logistic map within this method in order to clarify the effect 
of precision of $a_c$ values. To illustrate this point, in Fig.~2a we give 
three different cases for the logistic map. It is clearly seen that the 
$f(\alpha)$ curve obtained using high precision for $a_c$ (circle) does not 
coincide with the one obtained using less precision for $a_c$ (dotted line) 
for the same number of iterations ($I=1024$). On the other hand, we realize 
that it is possible to obtain the correct curve with less precision in 
$a_c$ if smaller values of iteration numbers are used ($I=256$) as it is 
evident from the Fig.~2a (full line). 
The same kind of behaviour for the $f(\alpha)$ curve has been observed for 
the Henon map as shown in Fig.~2b. Analyzing the results given in this figure, 
one can easily conclude that for the Henon map ($b\neq 0$) the precision 
of the $a_c$ value used in calculations is not enough when larger number of 
iterations ($I=1024$) is used, whereas the curve coincides with that of the 
logistic case (as expected) if smaller number of iterations is used ($I=256$). 
It is worth mentioning that the correct value of box counting fractal 
dimension $d_f$ of the critical attractor of the Henon map, which is the 
maximum value of the $f(\alpha)$ curve, is consistent with the above 
discussion (see Fig.~2). 
Therefore, within the light of this analysis, these results are quite 
convincing to conclude that the $q^*$ value of the Henon map which comes from 
the second method is the same as the one obtained from the first method. \\

\noindent {\bf Third Method -} ~ Finally, we shall use the third method 
(i.e., entropy increase rate procedure) to estimate the proper $q^*$ value 
of the Henon map in order to strengthen the results of the first and the 
second methods. To do this, we implement the following 
procedure \cite{latbar2,ugur} : firstly, we partition the phase space into 
$W$ equal cells ($W=1000\times 1000$), then we choose one of these cells and 
select $N$ initial conditions all inside this cell. As time evolves, these 
$N$ points spread within the phase space, yielding a set of probabilities. 
In the beginning of time, naturally $S_q(0)=0$, then the entropy increase 
rate gradually exhibits three successive regions. 
The entropy is almost constant in time in the first region, then it starts 
increasing in the second 
(intermediate) region and finally saturates in the third one. Among 
these regions, the intermediate one is the region where the linear increase 
of the proper entropy is expected to emerge \cite{latbar2,ugur}. 
Time evolution of $S_q(t)$ for the Henon map is given in Fig.~3. 
It is seen that, in the intermediate region, the linear increase of the 
entropy with time emerges only for a special value of $q$, and this value 
corresponds to the $q^*$ value (within a good precision) determined 
previously from the other two methods. 
When $q\neq q^*$, the entropy curves upwards (if $q<q^*$) or downwards 
(if $q>q^*$). In order to quantitatively support this picture, we fit the 
curves with the polynomial $S_q(t)= A+Bt+Ct^2$ in the intermediate region 
[$t_1,t_2$]. Since the nonlinearity coefficient $R\equiv C(t_1+t_2)/B$ is a 
measure of the importance of the nonlinear term, it should vanish for a 
strictly linear fit. In the inset of Fig.~3, we present the behaviour of $R$ 
from where we estimate the proper value of $q$ as $q^*=0.25$, which is the 
value of the $q$ index when $R$ is strictly zero.\\

Summing up, we analyzed the mixing properties of a two-dimensional dissipative 
map at the edge of chaos using three distinct methods that are developed and 
tested so far for many one-dimensional map families. As being the first example 
that numerically verifies the scenario in two-dimensional maps, this work would 
hopefully stimulate new projects addressing more general cases like 
high dimensional dissipative and/or conservative dynamical systems.  
Along these lines, in an ongoing project which will be reported elsewhere, 
we analyze another two-dimensional map, the Lozi-type map \cite{aziz}, to 
provide additional support for the present scenario.

\section*{Acknowledgments}
Financial support of Turkish Academy of Sciences from the 
TUBA/GEBIP Program is acknowledged. I also thank Dr. Aziz-Alaoui for 
his invaluable help on the calculation of the Lyapunov exponents.



\vspace{1.5cm}

{\bf Figure Captions}

\vspace{1.5cm}

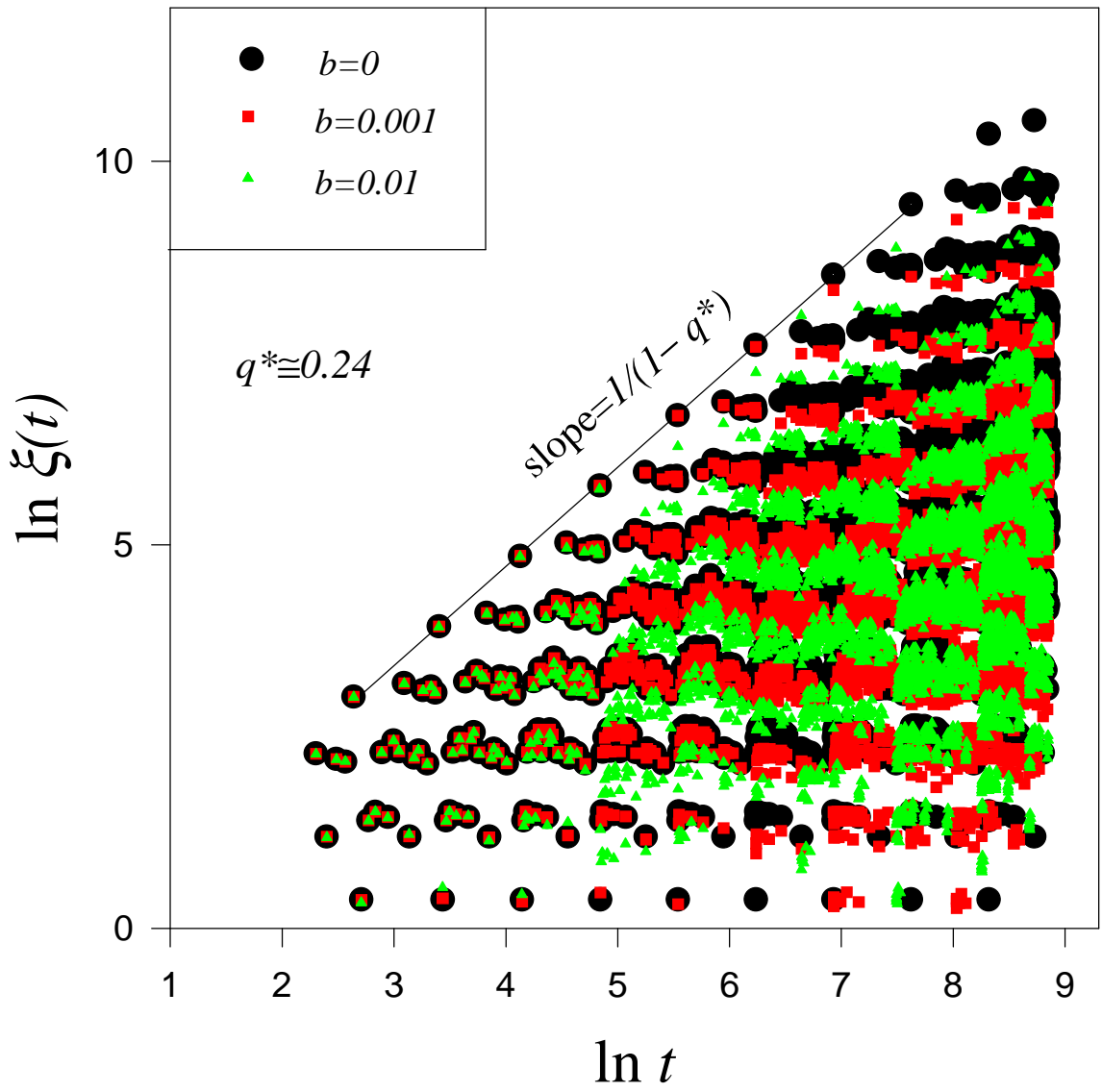
{\bf Figure 1} - Time evolution of the sensitivity function in a log-log 
plot for various values of $b$ parameter.\\

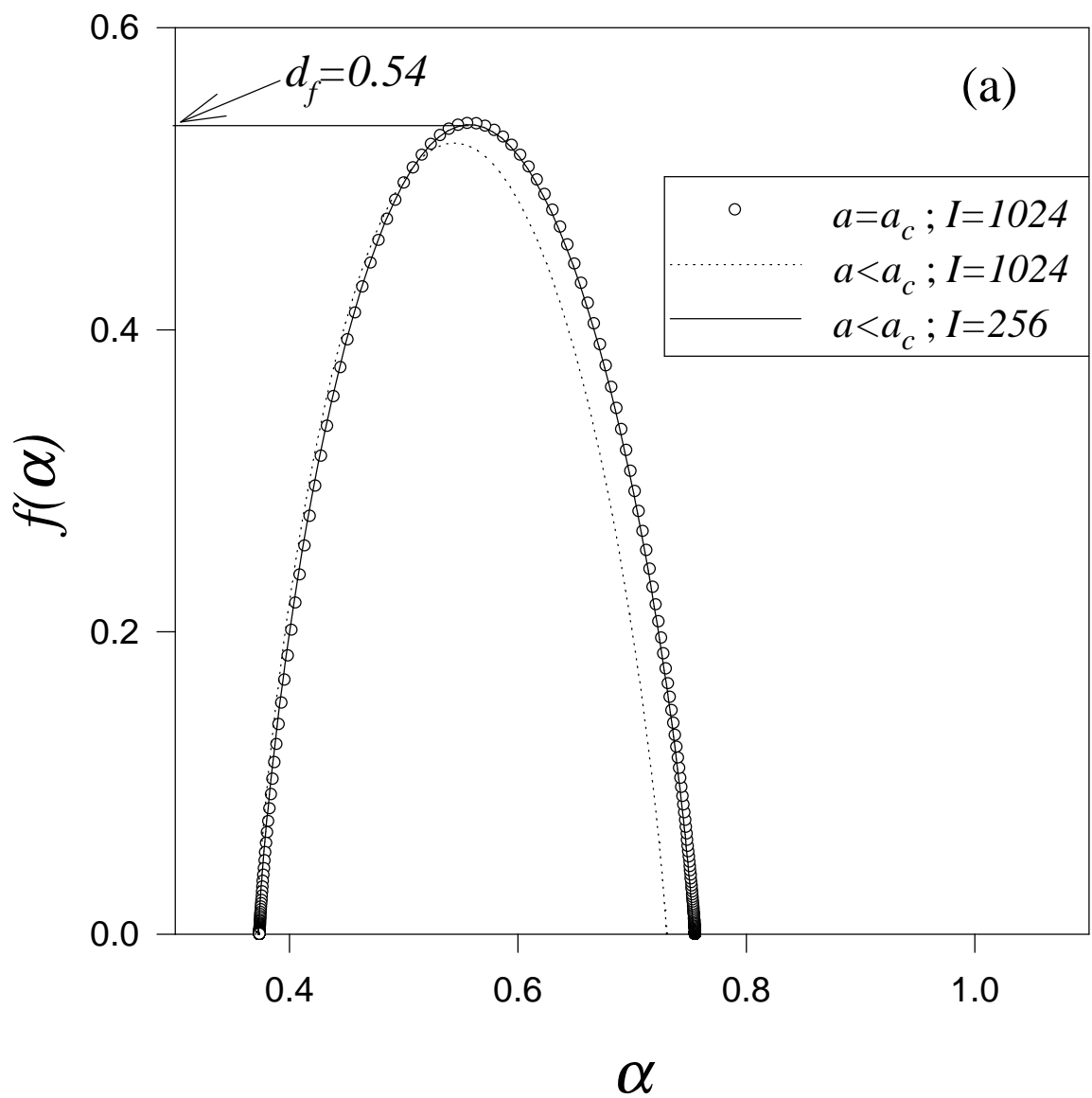
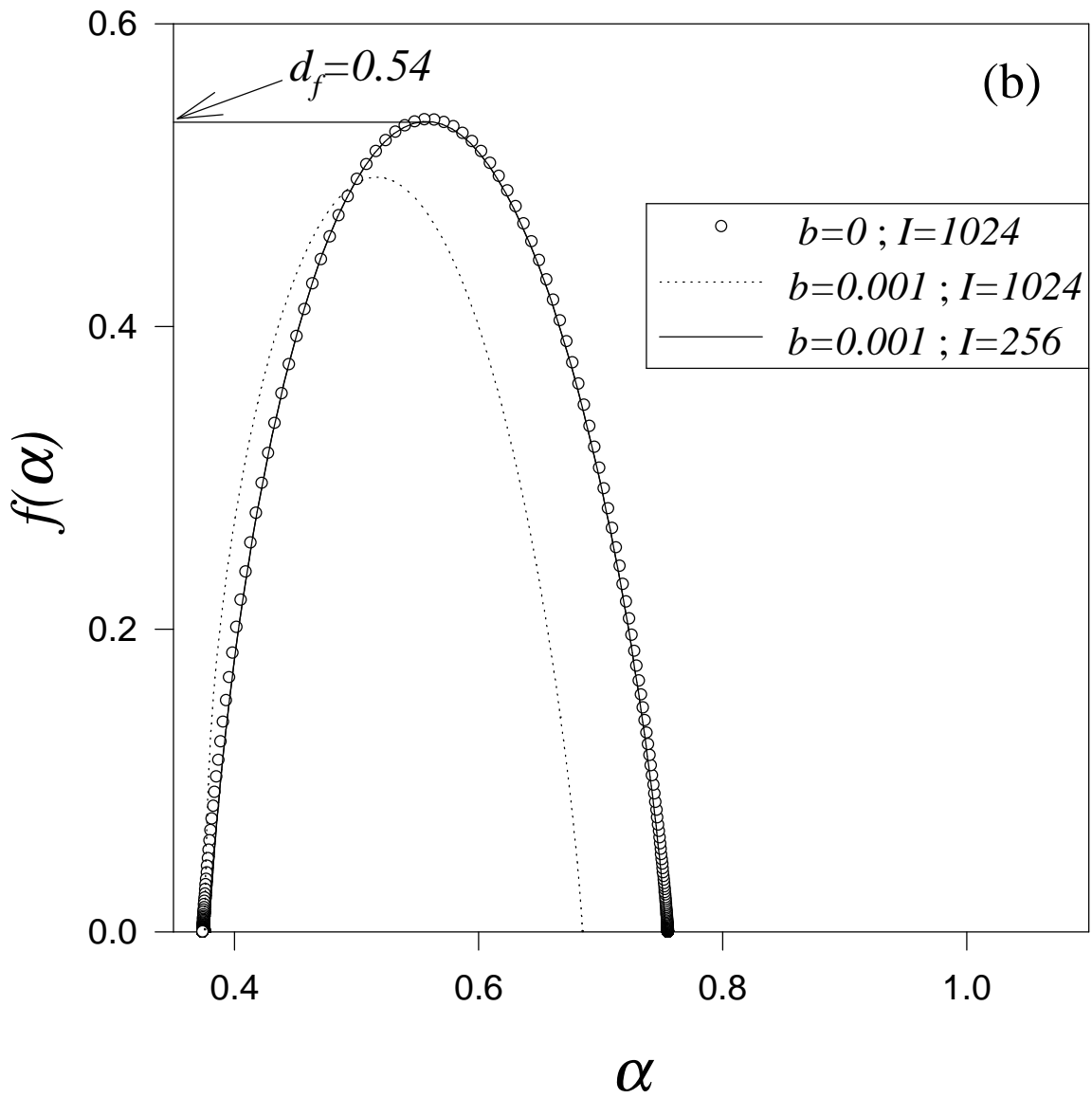
{\bf Figure 2} - The behaviour of $f(\alpha)$ curve (a) for the logistic 
map, (b) for the Henon map. \\

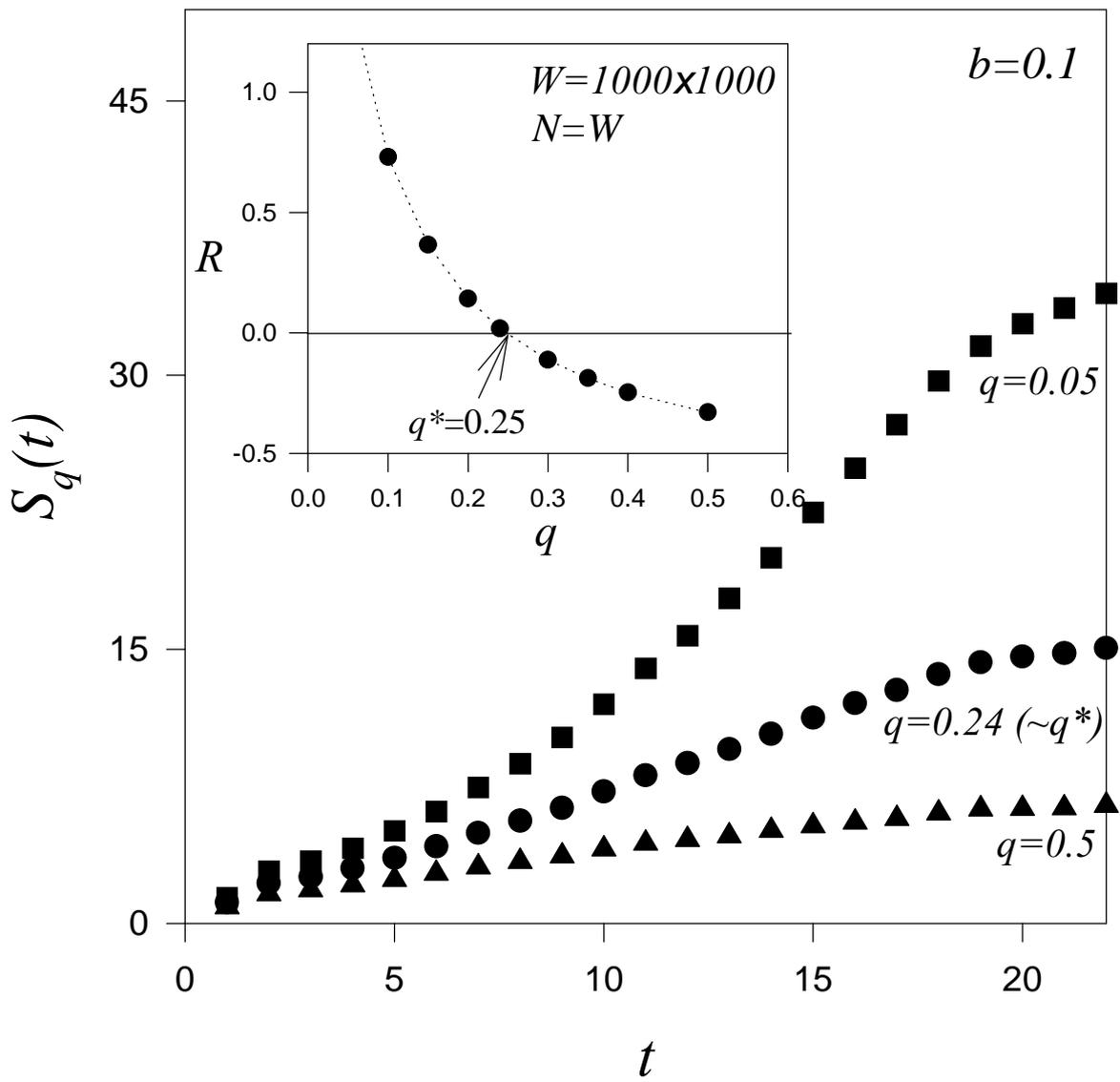
{\bf Figure 3} - Time evolution of the Tsallis entropy for $b=0.1$ for three 
different values of $q$. Inset: The nonlinearity coefficient $R$ versus $q$. 
The interval characterizing the intermediate region is [7,19]. The dotted 
lines is guide to the eye.\\

\end{document}